\documentclass[twocolumn,showpacs,showkeys,preprintnumbers,amssymb,aps,superscriptaddress,prb]{revtex4}
\usepackage{graphicx}
\usepackage{dcolumn}
\usepackage{bm}

\usepackage{epsfig}

\begin{document}

\preprint{APS/123-QED}

\title{Intermediate magnetization plateaus in the spin-1/2 Ising-Heisenberg and Heisenberg models on two-dimensional triangulated lattices}

\author{Jana \v{C}is\'{a}rov\'{a}}
\affiliation{Department of Theoretical Physics and Astrophysics, Faculty of Science,
P. J. \v{S}af\'{a}rik University, Park Angelinum 9, 040 01, Ko\v{s}ice, Slovak republic}
\affiliation{Institute of Theoretical Physics, Ecole Polytechnique F\'{e}d\'{e}rale de Lausanne (EPFL), CH-1015 Lausanne, Switzerland}
\author{Fr\'ed\'eric Michaud}
\affiliation{Institute of Theoretical Physics, Ecole Polytechnique F\'{e}d\'{e}rale de Lausanne (EPFL), CH-1015 Lausanne, Switzerland}
\author{Fr\'ed\'eric Mila}
\affiliation{Institute of Theoretical Physics, Ecole Polytechnique F\'{e}d\'{e}rale de Lausanne (EPFL), CH-1015 Lausanne, Switzerland}
\author{Jozef Stre\v{c}ka}
\affiliation{Department of Theoretical Physics and Astrophysics, Faculty of Science,
P. J. \v{S}af\'{a}rik University, Park Angelinum 9, 040 01, Ko\v{s}ice, Slovak republic}

\date{\today}

\begin{abstract}
The ground state and zero-temperature magnetization process of the spin-1/2 Ising-Heisenberg model on two-dimensional triangles-in-triangles lattices is exactly calculated using eigenstates of the smallest commuting spin clusters. Our ground-state analysis of the investigated classical--quantum spin model reveals three unconventional dimerized or trimerized quantum ground states besides two classical ground states. It is demonstrated that the spin frustration is responsible for a variety of magnetization scenarios with up to three or four intermediate magnetization plateaus of either quantum or classical nature. The exact analytical results for the Ising-Heisenberg model are confronted with the corresponding results for the purely quantum Heisenberg model, which were obtained by numerical exact diagonalizations based on the Lanczos algorithm for finite-size spin clusters of 24 and 21 sites, respectively. It is shown that the zero-temperature magnetization process of both models is quite reminiscent and hence, one may obtain some insight into the ground states of the quantum Heisenberg model from the rigorous results for the Ising-Heisenberg model even though exact ground states for the Ising-Heisenberg model do not represent true ground states for the pure quantum Heisenberg model.
\end{abstract}



\maketitle

\section{Introduction}

Two-dimensional frustrated quantum spin systems have attracted a great amount of research interest over the last few decades.\cite{lacr11} Frustration as a phenomenon originating from competing interactions incompatible with a lattice geometry prevents spins from simultaneously satisfying all pair spin-spin interactions and may thus lead to a variety of fascinating quantum phenomena. Among the most interesting properties emerging in frustrated quantum spin systems one could mention the existence of exotic ground states with a spin-liquid or valence-bond-solid character.\cite{lacr11,lhui02,misg04,bale10} At sufficiently low temperatures, the external magnetic field may additionally cause the appearance of intermediate magnetization plateaus at some fractional values of the saturation magnetization.\cite{totsuka,mila,rich04,hone04,schu04,takigawa_mila}

SrCu$_2$(BO$_3$)$_2$ represents one of the most famous experimental realizations of the frustrated quantum spin systems, which display an intriguing sequence of magnetization plateaus before reaching the saturation magnetization. High-field measurements performed on this layered magnet have revealed intermediate plateaus at $\frac{1}{8}$, $\frac{1}{4}$, and $\frac{1}{3}$ of the saturation magnetization.\cite{kage99,kage00,kage02} Subsequently, torque measurements by Sebastian \textit{et al}.\cite{seba08} have suggested the presence of additional striking fractional plateaus. A similar conclusion has also been reached more recently on the basis of torque and NMR measurements\cite{takigawa}, with a different sequence of plateaus however. From the theoretical point of view,
deep insight into the low-temperature magnetization process of SrCu$_2$(BO$_3$)$_2$ has been provided by the investigation of the spin-$\frac{1}{2}$ quantum Heisenberg model on the Shastry-Sutherland lattice\cite{miya03,dori08,aben08,isae09}, even if the actual sequence of plateaus for large
inter-dimer coupling is still debated.

The existence of an intermediate magnetization plateau at one third of the saturation magnetization has been recently reported for the family of polymeric coordination compounds Cu$_9$X$_2$(cpa)$_6$$\cdot$nH$_2$O (X=F,Cl,Br and cpa=carboxypentonic acid).\cite{gonz93,maru94,atec95,meka98,okub98,meka01} It provides a remarkable experimental realization of the spin-$\frac{1}{2}$ Heisenberg model on a quite exotic triangulated kagom\'e lattice.\cite{stre08,yao08,stre09,nato97,stre07,isod11,isod12,bati03,chen12} It actually turns out that the magnetic structure of this series of magnetic compounds is formed by two inequivalent lattice positions of the spin-$\frac{1}{2}$ Cu$^{2+}$ ions, whereas smaller equilateral triangles of Cu$^{2+}$ ions (Heisenberg trimers) are mutually inter-connected through Cu$^{2+}$ ions (Heisenberg monomers) situated at lattice points of a simple kagom\'e lattice. From the geometric point of view, the overall magnetic structure thus consists of smaller triangles of magnetic ions embedded in the larger triangles of a kagom\'e lattice and hence, it belongs to a class of two-dimensional triangles-in-triangles (TIT) lattices. Recent exact solution for the spin-$\frac{1}{2}$ Ising-Heisenberg model on two closely related TIT lattices descending from a simple triangular lattice rather than from a kagom\'e lattice gave evidence of a significant impact of local quantum fluctuations on the overall magnetic behavior of this hybrid classical--quantum spin model.\cite{cisa12} The main purpose of this work is to find possible ground states of the spin-$\frac{1}{2}$ Ising-Heisenberg model in the presence of an external magnetic field and to compare its zero-temperature magnetization process with that of the full quantum version of the spin-$\frac{1}{2}$ Heisenberg model.

The outline of this paper is as follows. In Sec. \ref{sec:model} we introduce the spin-$\frac{1}{2}$ Ising-Heisenberg and Heisenberg models on two TIT lattices and briefly describe the procedure used for the calculation of the zero-temperature magnetization process. The most interesting results for the ground state and zero-temperature magnetization curves are presented in Sec. \ref{sec:result}. Finally, the summary of the most important scientific achievements and future outlooks are mentioned in Sec. \ref{sec:conclusion}.

\section{Model and its ground state}
\label{sec:model}

\begin{figure}[t]
\vspace{0.0cm}
\includegraphics[width=6.0cm]{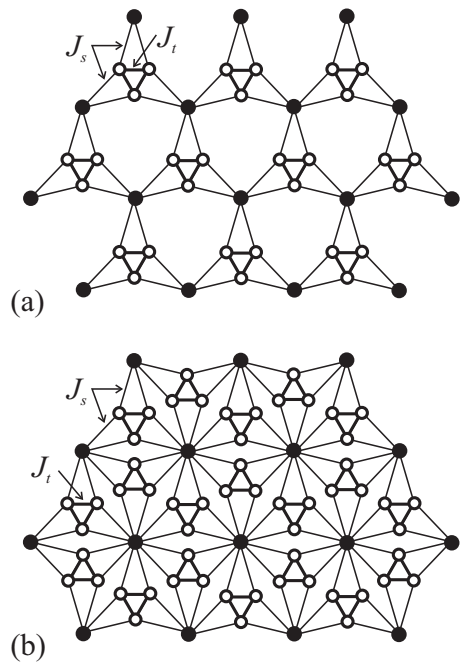}
\vspace{-0.2cm}
\caption{Two particular examples of the TIT lattices derived from a simple triangular lattice. The lattice positions solely belonging to
the inner equilateral (outer isosceles) triangles are schematically represented by empty (filled) circles.}
\label{fig1}
\end{figure}

\begin{figure}[t]
\vspace{0.0cm}
\includegraphics[width=3.0cm]{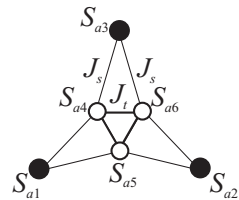}
\vspace{-0.2cm}
\caption{Elementary spin cluster in the form of a spin star.}
\label{fig2}
\end{figure}

Let us introduce first the underlying magnetic lattice for frustrated quantum spin models, whose zero-temperature magnetization process will be explored in detail. Fig.~\ref{fig1} schematically illustrates two particular examples of the TIT lattices, which can be derived from a simple triangular lattice by placing an additional inner triangle either into each up-pointing triangle (Fig.~\ref{fig1}a) or into each triangle (Fig.~\ref{fig1}b) of a triangular lattice. The two displayed TIT lattices can alternatively be viewed as being composed of inter-connected \textit{stars} (see Fig.~\ref{fig2}), which have an equilateral triangle in their center and are joined together through outer isosceles triangles attached to each side of the inner equilateral triangles. Clearly, there are two inequivalent lattice sites in the TIT lattices, the ones forming the inner equilateral triangles shown by empty circles and the other ones solely belonging to the outer isosceles triangles depicted by filled circles. The more general anisotropic version of the spin-$\frac{1}{2}$ Heisenberg model on two different but topologically related TIT lattices can be defined through the Hamiltonian
\begin{eqnarray}
\hat{\cal H} = J_t  \sum_{<i,j>}^{\circ\!-\!\circ} \hat{\textbf{S}}_i \cdot \hat{\textbf{S}}_j
+ J_s \sum_{<k,l>}^{\circ\!-\!\bullet} {(\hat{\textbf{S}}_k \cdot \hat{\textbf{S}}_l)}_{\Delta} - h \sum_{p} \! \! \hat{S}_p^z,
\label{Htot}
\end{eqnarray}
where $(\hat{\textbf{S}}_k \cdot \hat{\textbf{S}}_l)_{\Delta} = \Delta (\hat{S}_{k}^{x} \hat{S}_{l}^{x} + \hat{S}_{k}^{y} \hat{S}_{l}^{y}) + \hat{S}_{k}^{z} \hat{S}_{l}^{z}$ and $\hat{\textbf{S}}_{i} = (\hat{S}_{i}^{x}, \hat{S}_{i}^{y}, \hat{S}_{i}^{z})$ denotes the standard spin-$\frac{1}{2}$ operator. The interaction term $J_t$ describes the isotropic Heisenberg interaction between the nearest-neighbor spins from the inner equilateral triangles while the interaction term $J_s$ labels the anisotropic XXZ Heisenberg interaction between the nearest-neighbor spins from the inner equilateral and outer isosceles triangles, respectively. Finally, the last Zeeman's term $h$ accounts for the magnetostatic energy of all spins in an external magnetic field. In what follows, our main attention will be concentrated on two special cases of the Hamiltonian (\ref{Htot}) either with $\Delta = 0$ or $1$, respectively. While the Hamiltonian (\ref{Htot}) with $\Delta = 0$ reduces to the hybrid classical--quantum Ising-Heisenberg model,\cite{cisa12} the Hamiltonian (\ref{Htot}) with $\Delta = 1$ corresponds to the isotropic quantum Heisenberg model.

\subsection{Ising-Heisenberg model}

First, we turn our attention to the spin-$\frac{1}{2}$ Ising-Heisenberg model defined through the Hamiltonian (\ref{Htot}) with $\Delta = 0$. It is worthwhile to remark that the specific choice $\Delta = 0$ disregards the transverse part of the exchange interaction $J_s$, which consequently becomes the pure Ising coupling. Owing to this fact, the spins from the outer isosceles triangles are merely coupled to their nearest neighbors by the Ising interaction, while the isotropic Heisenberg interaction $J_t$ between the nearest-neighbor spins from the inner equilateral triangles is still preserved. For easy reference, the former spins from the outer isosceles triangles will be therefore referred to as the Ising spins and the latter spins from the inner equilateral triangles as the Heisenberg spins.

For further convenience, the total Hamiltonian (\ref{Htot}) can be decomposed into a sum of cluster Hamiltonians $\hat{\cal H}_a$
\begin{eqnarray}
\hat{\cal H} = \sum_{a=1}^{\gamma N} {\hat{\cal H}_a},
\label{Hsum}
\end{eqnarray}
where $N$  is the total number of Ising spins, $\gamma = 1$ ($\gamma = 2$) for the first (second) TIT lattice, so that $\gamma N $ is the total number of six-spin clusters schematically visualized in Fig.~\ref{fig2}. The cluster Hamiltonian $\hat{\cal H}_a$ is given by
\begin{eqnarray}
\hat{\cal H}_{a} \!\!\!&=&\!\!\! J_t \sum_{i=4}^{6} \hat{\textbf{S}}_{a,i} \cdot \hat{\textbf{S}}_{a,i+1} +
                   J_s \sum_{i=1}^{3} \!\! \left( \hat{S}_{a,i}^z \hat{S}_{a,i+3}^z + \hat{S}_{a,i}^z \hat{S}_{a,i+4}^z \! \right) \nonumber \\
                 \!\!\!&-&\!\!\! h \sum_{i=4}^6 \hat{S}_{a,i}^z - \frac{1}{3 \gamma} h \sum_{i=1}^3 \hat{S}_{a,i}^z,
\label{Hk}
\end{eqnarray}
with the convention $S_{a,7} \equiv S_{a,4}$.

It is worth noticing that the factor $\frac{1}{3 \gamma}$ emerging in the last term ensures correct counting of the Zeeman's energy of the Ising spins,
which is equally split into the $3 \gamma$ different cluster Hamiltonians involving a given Ising spin. Moreover, it is of fundamental importance
that different cluster Hamiltonians commute with each other $[\hat{\cal H}_i, \hat{\cal H}_j] = 0$ as they have in common at most one Ising spin.
Accordingly, different cluster Hamiltonians can be diagonalized independently of each other as they belong to mutually orthogonal Hilbert subspaces
and one consequently gets the factorized ground state as a tensor product over the lowest-energy eigenvectors of the cluster Hamiltonians (\ref{Hk}).
The minimization of all clusters simultaneously is not always possible if they share common spins, but in our case the different ground-state verify this condition.

In order to find the exact ground state of the spin-$\frac{1}{2}$ Ising-Heisenberg model on the two considered TIT lattices, it is thus sufficient to solve the eigenvalue problem for the cluster Hamiltonian (\ref{Hk}) by considering all possible configurations of three Ising spins involved therein.
For the sake of brevity, let us merely list all eigenvalues of the cluster Hamiltonian (\ref{Hk}) without quoting the corresponding eigenvectors, which will be later specified just for the eigenstates that may become the ground state. The eigenvectors of the cluster Hamiltonian can be classified by the total magnetization of the Ising spins, because this operator commutes with the Hamitonian.  The eigenvalues of the cluster Hamiltonian (\ref{Hk}) for three Ising spins equally aligned into the external-field direction $|S_{a,1}^z S_{a,2}^z S_{a,3}^z \rangle = |\!\!\uparrow \uparrow \uparrow \rangle$ read
\begin{eqnarray}
E_{1,2}^{\uparrow \uparrow \uparrow} &=& \frac{3}{4} J_t \mp \frac{3}{2} J_s \pm \frac{3}{2} h - \frac{1}{2 \gamma} h, \nonumber \\
E_{3,4}^{\uparrow \uparrow \uparrow} &=& - \frac{3}{4} J_t + \frac{1}{2} J_s - \frac{1}{2} h - \frac{1}{2 \gamma} h,\nonumber \\
E_{5,6}^{\uparrow \uparrow \uparrow} &=& -\frac{3}{4} J_t - \frac{1}{2} J_s + \frac{1}{2} h - \frac{1}{2 \gamma} h, \nonumber \\
E_{7,8}^{\uparrow \uparrow \uparrow} &=& \frac{3}{4} J_t \pm \frac{1}{2} J_s \mp \frac{1}{2} h - \frac{1}{2 \gamma} h.
\label{Euuu}
\end{eqnarray}
The energy spectrum of the cluster Hamiltonian (\ref{Hk}) for three Ising spins aligned opposite to the external magnetic field ($|S_{a,1}^z S_{a,2}^z S_{a,3}^z \rangle = |\!\!\downarrow \downarrow \downarrow \rangle$) can be obtained by changing the sign of the magnetic field $h$ in Eq.~(\ref{Euuu}).

If two Ising spins have the same orientation with respect to the external magnetic field and one Ising spins is in the opposite direction, for instance
$|S_{a,1}^z S_{a,2}^z S_{a,3}^z \rangle = |\!\!\uparrow \uparrow \downarrow \rangle$, one obtains the following eigenvalues of the cluster Hamiltonian (\ref{Hk})
\begin{eqnarray}
E_{1,2}^{\uparrow \uparrow \downarrow} &=& \frac{3}{4} J_t \pm \frac{1}{2} J_s \mp \frac{3}{2} h - \frac{1}{6\gamma} h, \nonumber \\
E_{3,4}^{\uparrow \uparrow \downarrow} &=& - \frac{3}{4} J_t \pm \frac{1}{2} J_s \mp \frac{1}{2} h - \frac{1}{6\gamma} h, \nonumber \\
E_{5,6}^{\uparrow \uparrow \downarrow} &=& \pm \frac{1}{2} B^{+} - \frac{1}{2} h - \frac{1}{6\gamma} h, \nonumber \\
E_{7,8}^{\uparrow \uparrow \downarrow} &=& \pm \frac{1}{2} B^{-} + \frac{1}{2} h - \frac{1}{6\gamma} h,
\label{Euud}
\end{eqnarray}
where
\begin{equation}
B^{\pm} = \sqrt{\left( \frac{1}{2} J_t \pm J_s \right)^2 + 2 J_t^2}.
\label{Bpm}
\end{equation}

We can get the energy spectrum of the cluster Hamiltonian (\ref{Hk}) for one Ising spin pointing in the direction of the external magnetic field
and two Ising spins aligned in the opposite direction $|S_{a,1}^z S_{a,2}^z S_{a,3}^z \rangle = |\!\!\uparrow \downarrow \downarrow \rangle$ by changing the sign of the magnetic field $h$ in Eq.~(\ref{Euud}).

The lowest-energy eigenvalue from the set given by Eqs.~(\ref{Euuu})-(\ref{Euud}) unambiguously determines the ground state of the Ising-Heisenberg model,
which will be investigated in detail in Sec. \ref{sec:result} depending on the relative strength of both interaction constants and the magnetic field. In addition, the zero-temperature magnetization can easily be obtained from the ground-state energy using the relation
\begin{eqnarray}
\frac{m}{m_s} = - \frac{2 \gamma}{3 \gamma + 1} \frac{\partial E_g}{\partial h},
\label{mag}
\end{eqnarray}
where the factor $\frac{2 \gamma}{3 \gamma + 1}$ ensures the proper normalization of the total magnetization with respect to its saturation value since the elementary unit cell consists of one Ising spin and three (six) Heisenberg spins for the first (second) TIT lattice displayed in Fig.~\ref{fig1}a (Fig.~\ref{fig1}b).

\subsection{Heisenberg model}

\begin{figure}[t]
\includegraphics[width=7.0cm]{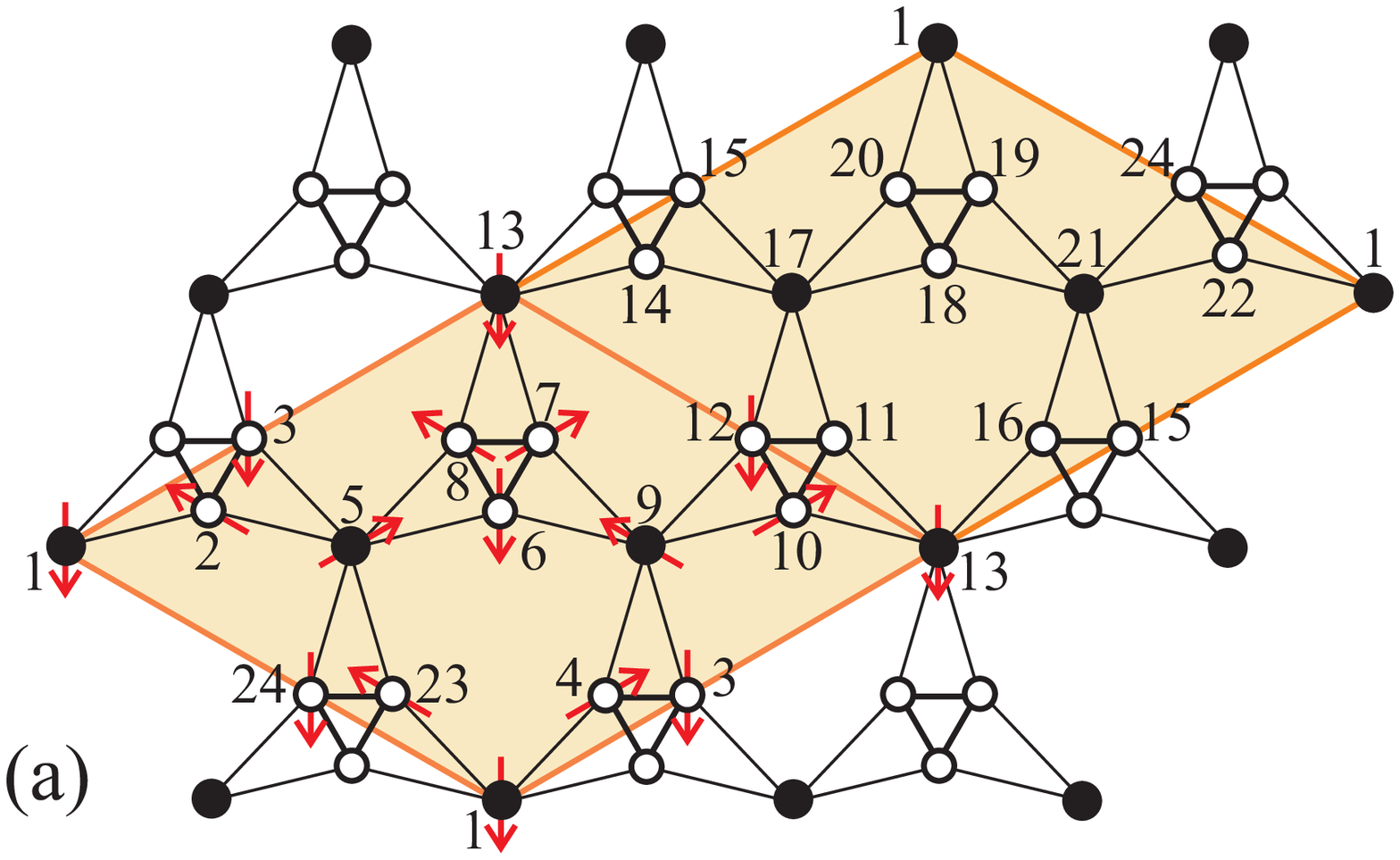}
\includegraphics[width=7.0cm]{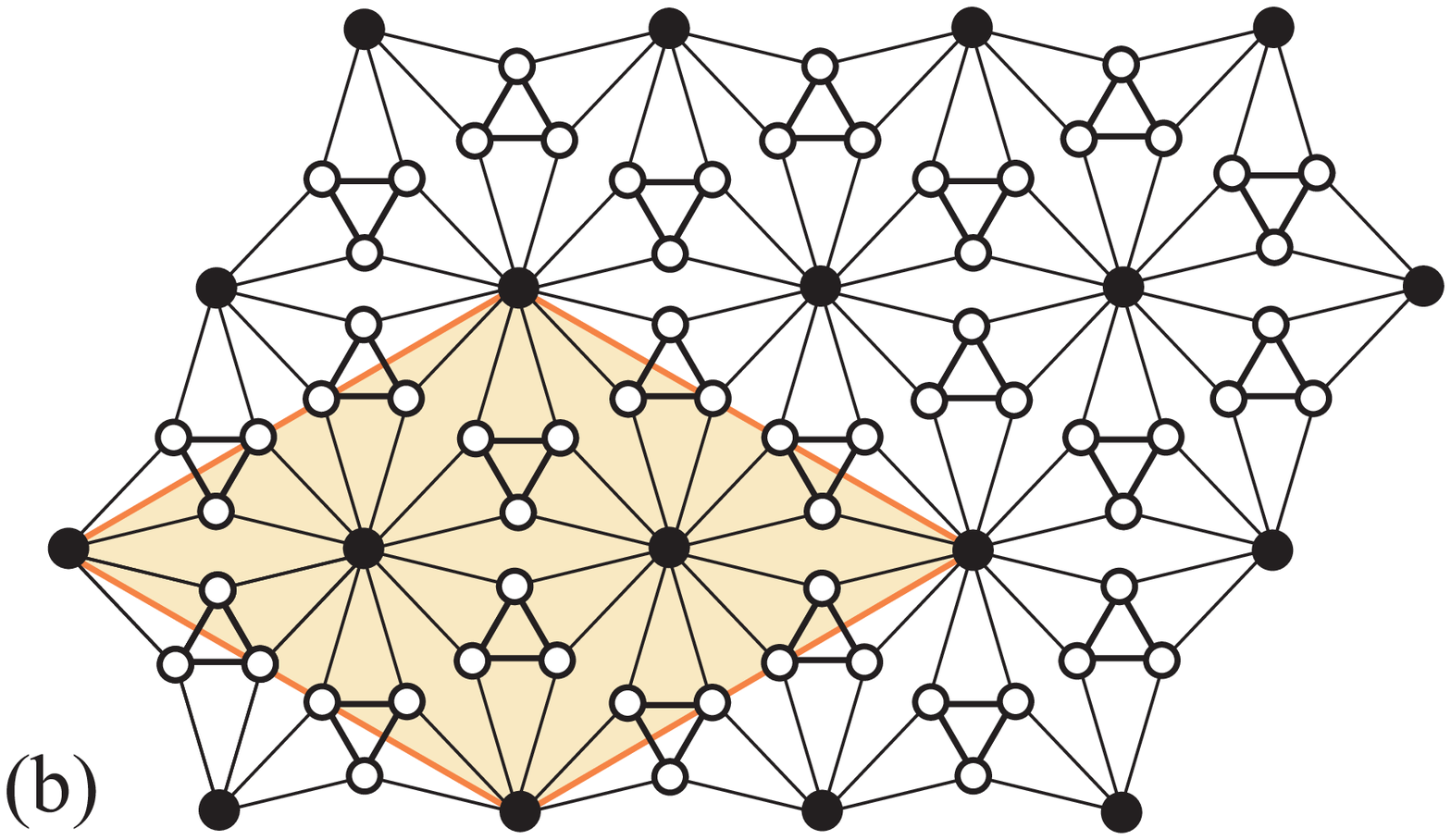}
\vspace{-0.2cm}
\caption{(a) The magnetic unit cell of the first TIT lattice is shown as the 12-site rhombus spin cluster delimited by thick (orange) line, which also includes spin orientations within the classical ground state. The 24-site parallelogram spin cluster constituted by two magnetic unit cells was used  for the exact diagonalization as the largest spin cluster; (b) The magnetic unit cell for the second TIT lattice is formed by the 21-site rhombus spin cluster, which was used for the exact diagonalization as the largest spin cluster (spin orientations within the classical ground state are not shown for clarity).}
\label{figed}
\end{figure}

The other particular case of the Hamiltonian (\ref{Htot}) with $\Delta = 1$ corresponds to the isotropic spin-$\frac{1}{2}$ Heisenberg model in a presence of the external magnetic field. Of course, the ground state of the fully quantum Heisenberg model on the TIT lattices cannot be rigorously obtained by analytical calculations and one has to rely on numerical methods. The lowest-energy eigenstates of the spin-$\frac{1}{2}$ Heisenberg model on two TIT lattices were obtained by employing the numerical exact diagonalization based on the Lanczos algorithm. Our numerical calculations were limited to relatively small finite-size subsystems when comparing them with the relevant magnetic unit cell of the classical ground state of both TIT lattices, which extends over three lattice unit cells. The magnetic unit cell of the classical ground state for the first TIT lattice thus contains 12 spins (the elementary unit cell consists of 4 spins), while our exact numerical diagonalization was limited to the rhombus and parallelogram spin clusters of 12 and 24 sites as specified in Fig.~\ref{figed}a. On the other hand, the magnetic unit cell of the classical ground state for the second TIT lattice already contains 21 spins (the elementary unit cell consists of 7 spins), which was also an upper limit for the spin-cluster size of our numerical diagonalization (see Fig.~\ref{figed}b).

Obviously, the total spin operator $\hat{S}^z_{tot} = \sum_{p} \hat{S}^z_p$ commutes with the Hamiltonian $[\hat{\cal H},\hat{S}^z_{tot}] = 0$, which means that the external field does not alter the relevant eigenstates and the total magnetization $S^z_{tot} = \sum_{p} S^z_p$ is a conserved quantity. Hence, the full energy spectrum of the spin-$\frac{1}{2}$ Heisenberg model in the presence of the external magnetic field can be obtained from the energy spectrum in a zero magnetic field according to
\begin{eqnarray}
E_i (h, S_{tot}^z) = E_i (0, S_{tot}^z) - h S_{tot}^z.
\label{m_sztot}
\end{eqnarray}
It is quite clear from Eq.~(\ref{m_sztot}) that the ground state may correspond only to the lowest-energy eigenstate from some sector with the fixed value of $S_{tot}^z$. The field-induced change in the ground-state spin arrangement can thus be related to crossings between the lowest-energy levels from the sectors with different $S_{tot}^z$. Hence, the zero-temperature magnetization curves can simply be constructed from the lowest-energy eigenstates, which were obtained with the help of numerical exact diagonalization based on the Lanczos algorithm for all possible sectors of $S^z_{tot}$ at zero magnetic field.

\section{Results and discussion}
\label{sec:result}

Before proceeding to the discussion of the most interesting results for the zero-temperature magnetization curves of the spin-$\frac{1}{2}$ Ising-Heisenberg and Heisenberg model on the TIT lattices, let us introduce a more convenient parametrization of the interaction constants
\begin{eqnarray}
J_t = J \sin \alpha, \qquad J_s = J \cos \alpha,  \qquad \alpha \in \left\langle 0, 90^{\circ} \right\rangle,
\label{jtjs}
\end{eqnarray}
which is useful in an attempt to provide a comprehensive analysis of the full parameter space. The specific choice $\alpha = 0$ corresponds to $J_t = 0$ and $J_s = J$, which means that the isotropic Heisenberg interaction along the inner equilateral triangles vanishes and the nearest-neighbor spins from the inner equilateral and outer isosceles triangles are merely coupled through the interaction $J_s = J$. On the other hand, the opposite limit $\alpha = 90^{\circ}$ implies that the investigated spin systems break into a set of independent Heisenberg trimers and monomers since $J_t = J$ and $J_s = 0$. For simplicity, our subsequent analysis will be restricted just to the most interesting particular case with the antiferromagnetic interactions $J_t>0$ and $J_s>0$, whose relative strength is determined by the newly introduced parameter $\alpha$ through $J_t/J_s = \tan \alpha$. Last but not least, it should be pointed out that the antiferromagnetic Heisenberg interaction $J_t$ is responsible for a spin frustration, which becomes relevant for sufficiently high values of the interaction ratio $J_t/J_s$ (or $\alpha$).

\begin{figure}[t]
\vspace{0.0cm}
\includegraphics[width=10cm]{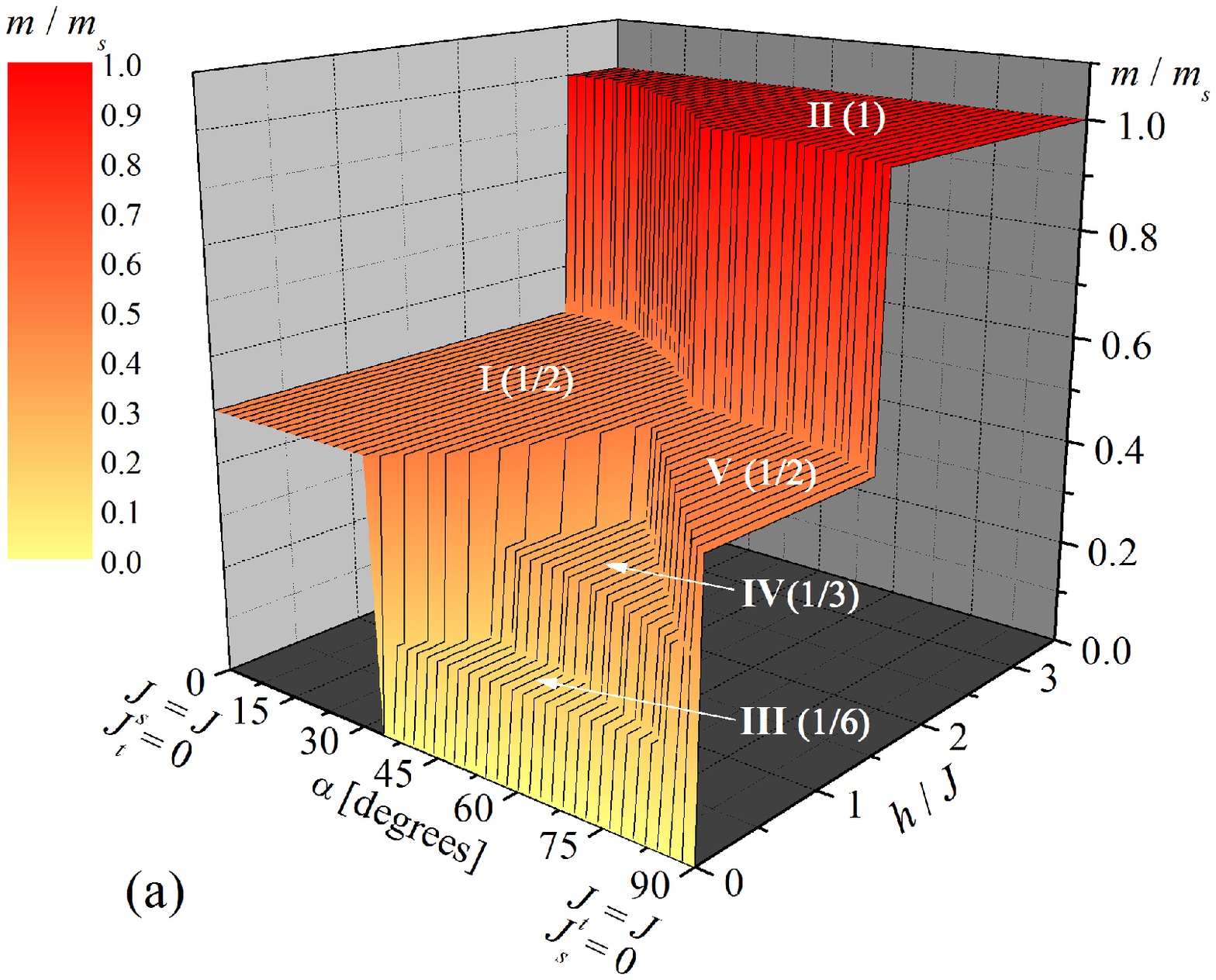}
\includegraphics[width=10cm]{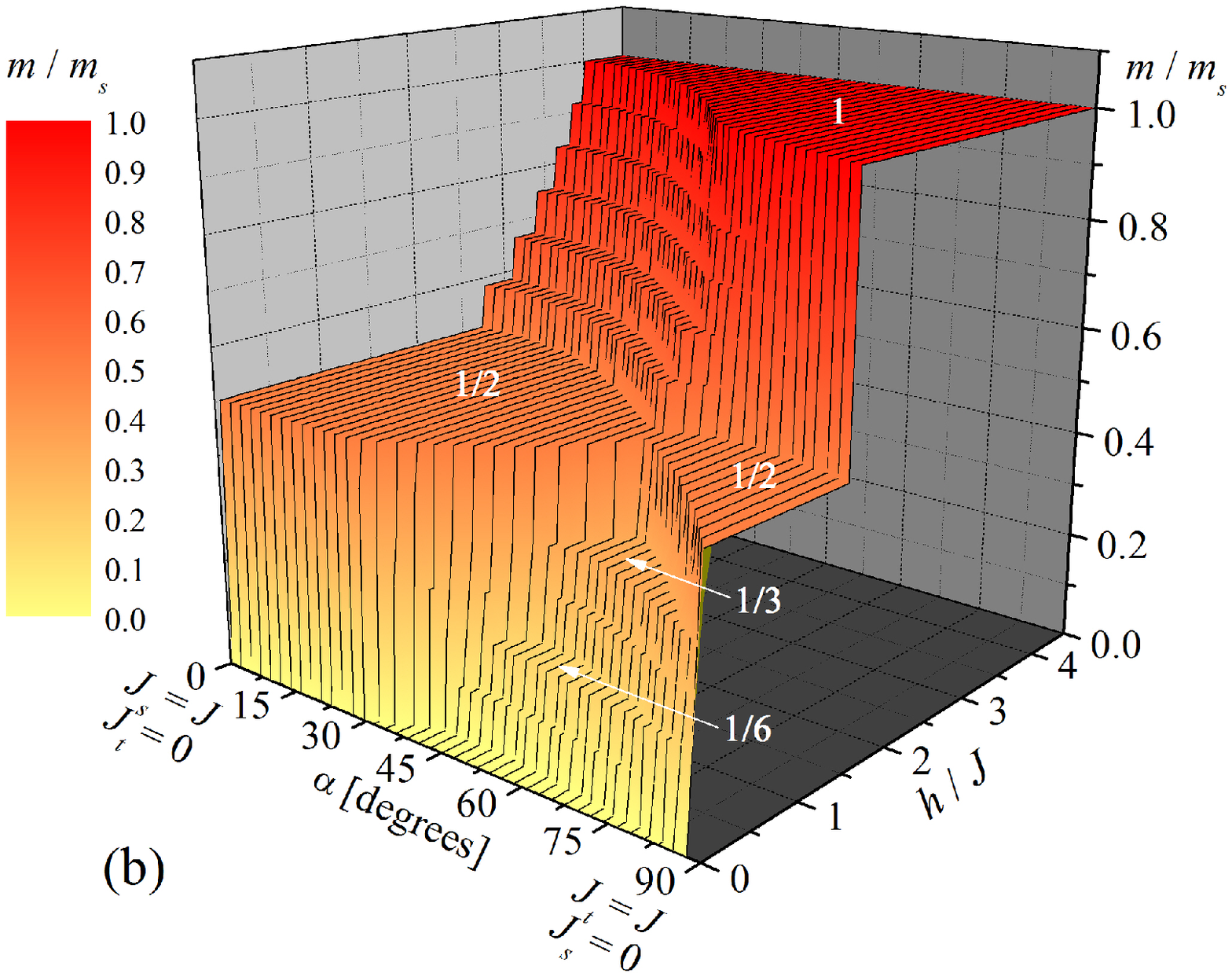}
\vspace{-0.7cm}
\caption{(a) A three-dimensional plot for the zero-temperature magnetization of the spin-$\frac{1}{2}$ Ising-Heisenberg model on the first TIT lattice (Fig.~\ref{fig1}a) as a function of the magnetic field $H/J$ and the parameter $\alpha$; (b) The same three-dimensional plot but for the pure quantum spin-$\frac{1}{2}$ Heisenberg model on the 24-site parallelogram spin cluster depicted in Fig.~\ref{figed}a.}
\label{fig3}
\end{figure}

\begin{figure*}[t]
\begin{center}
\includegraphics[width=5.0cm]{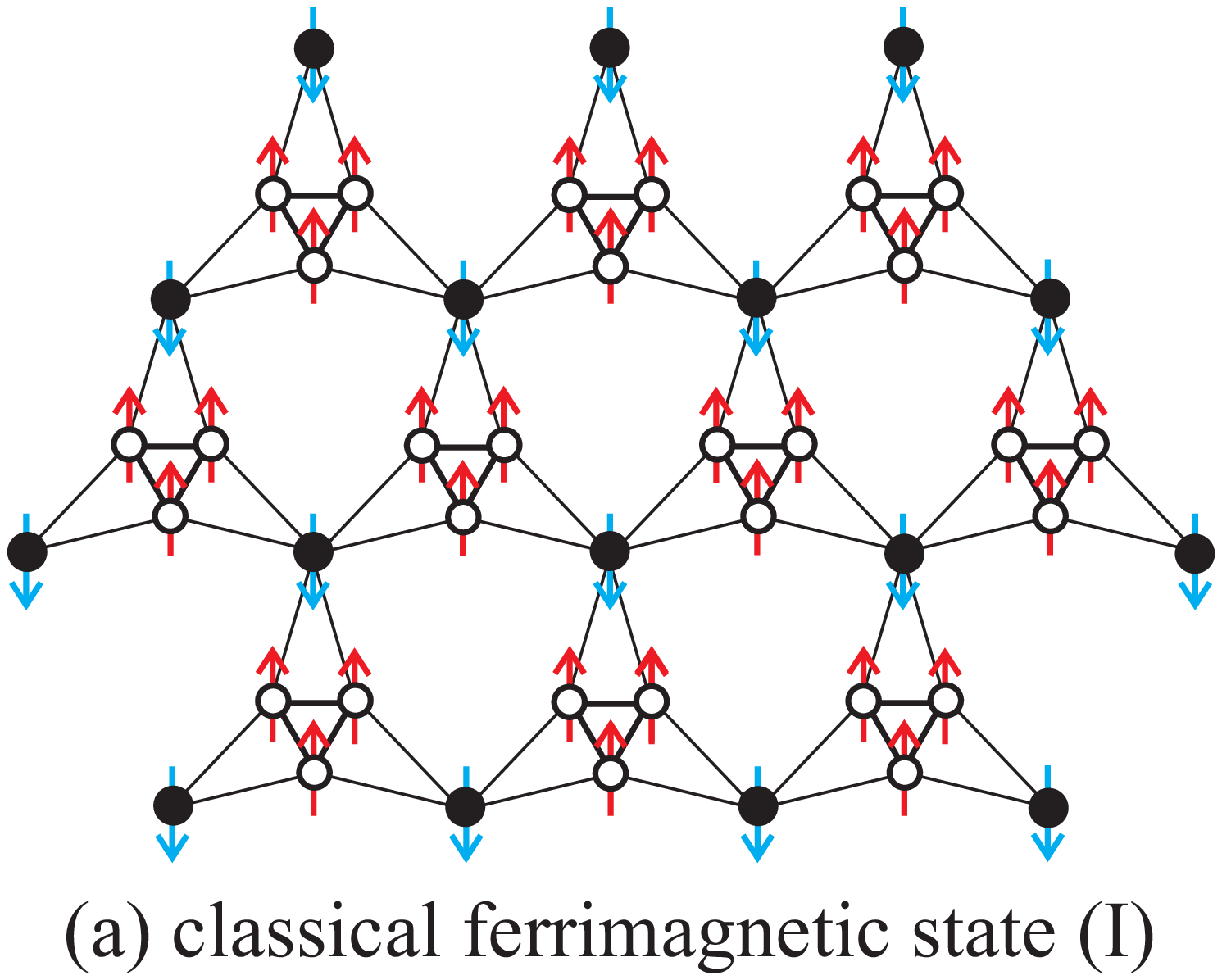}
\hspace{0.5cm}
\includegraphics[width=5.0cm]{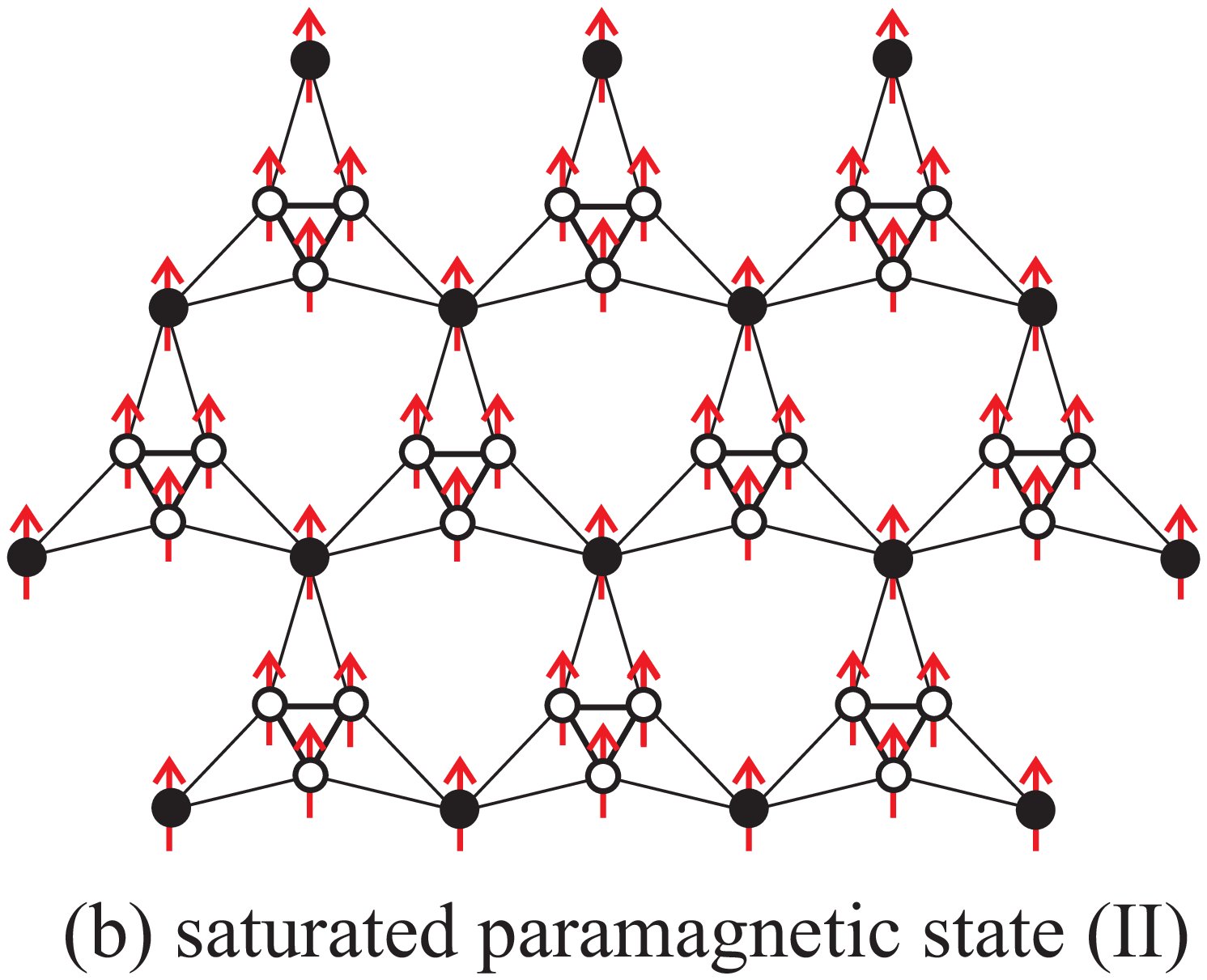}
\hspace{0.5cm}
\includegraphics[width=5.0cm]{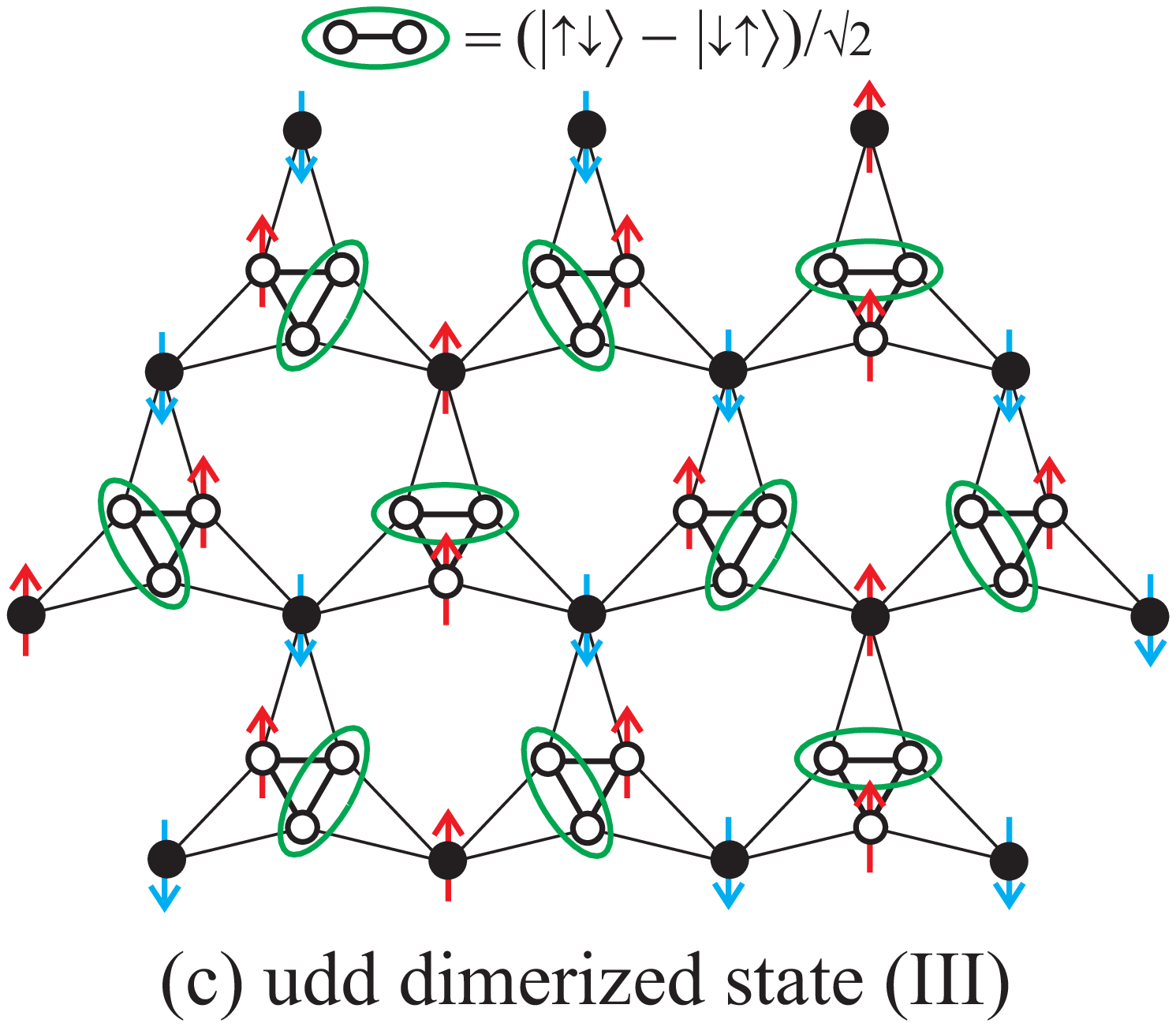}
\\
\vspace{0.3cm}
\includegraphics[width=5.0cm]{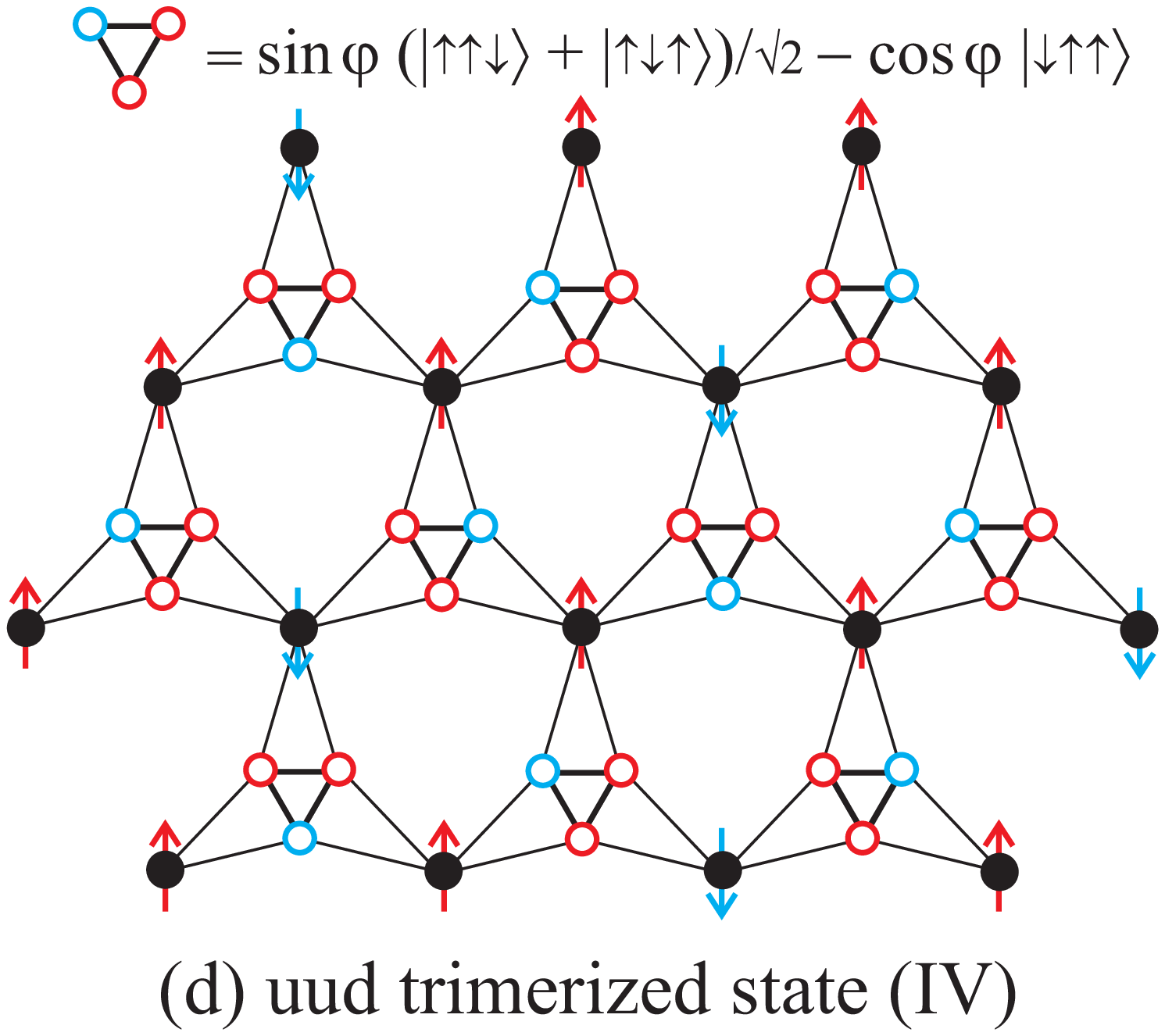}
\hspace{0.5cm}
\includegraphics[width=5.0cm]{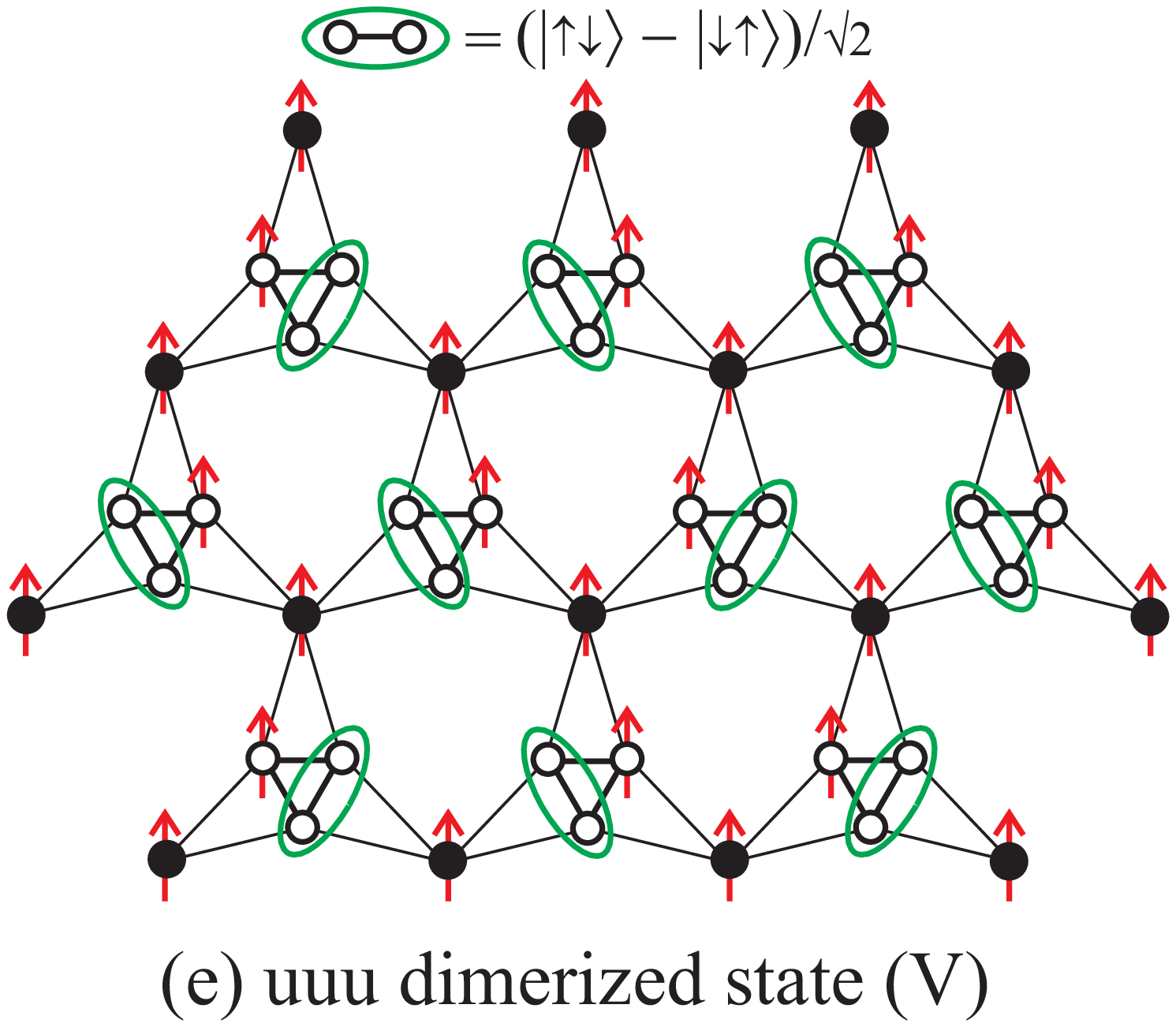}
\end{center}
\vspace{-0.6cm}
\caption{Two classical and three quantum ground states of the spin-$\frac{1}{2}$ Ising-Heisenberg model referred to as: (a) the classical ferrimagnetic state I;
(b) the saturated paramagnetic state II; (c) the uud dimerized state III; (d) the udd trimerized state IV; (e) the uuu dimerized state V.}
\label{fig4}
\end{figure*}

For the sake of comparison, the zero-temperature magnetizations of the spin-$\frac{1}{2}$ Ising-Heisenberg and Heisenberg model on the first TIT lattice are plotted in Fig.~\ref{fig3} against the magnetic field $H/J$ and the parameter $\alpha$. First, let us make a few comments on the magnetization curves of the spin-$\frac{1}{2}$ Ising-Heisenberg model for which the respective ground states can be found quite rigorously. As one can see from Fig.~\ref{fig3}a, the Ising-Heisenberg model displays five possible ground states two of which are classical and three quantum in character. If the interaction ratio $J_t/J_s<\frac{2}{3}$ (i.e. $\alpha<33.7^{\circ}$), the \textit{classical ferrimagnetic state} emerges in the ground state at low enough fields as a result of the following lowest-energy eigenstate of the spin star (see Fig.~\ref{fig2} for the definition of the sites inside a star)
\begin{eqnarray}
|{\rm I} \rangle &=& |\!\downarrow \downarrow \downarrow \rangle_{a1,a2,a3} \otimes |\!\uparrow \uparrow \uparrow\rangle_{a4,a5,a6}, \nonumber \\
E_{\rm I} &=& \frac{3}{4} J_t - \frac{3}{2} J_s - \frac{3}{2} h + \frac{1}{2 \gamma} h,
\label{I}
\end{eqnarray}
which indicates that the Heisenberg spins are fully aligned towards the magnetic field in contrast to the Ising spins pointing in an opposite direction (see Fig.~\ref{fig4}a for a schematic illustration of this classical spin arrangement). The total magnetization of the classical ferrimagnetic state will be consequently equal to one half of the saturation magnetization, because the magnetization of one Ising spin just partially compensates the magnetization of three Heisenberg spins per elementary unit cell of the first TIT lattice. The classical ferrimagnetic state ends up at the saturation field $h_{s} = 3 J_s$, above which the \textit{saturated paramagnetic state} with a perfect alignment of all spins into the external magnetic field emerges owing to the lowest-energy eigenstate (see Fig.~\ref{fig4}b)
\begin{eqnarray}
|{\rm II} \rangle &=& |\!\uparrow \uparrow \uparrow \rangle_{a1,a2,a3} \otimes |\!\uparrow \uparrow \uparrow \rangle_{a4,a5,a6}, \nonumber \\
E_{\rm II} &=& \frac{3}{4} J_t + \frac{3}{2} J_s - \frac{3}{2} h - \frac{1}{2 \gamma} h.
\label{II}
\end{eqnarray}
As could be expected, this classical mechanism for the formation of an intermediate magnetization plateau at half of the saturation magnetization comes from the dominant character of the Ising interaction $J_s$ that governs the magnetic behavior in the relevant parameter region.

A more striking situation appears for the other possible choice of the interaction ratio $J_t/J_s>\frac{2}{3}$ (i.e. $\alpha>33.7^{\circ}$), which promotes spin frustration and the gradual strengthening of local quantum fluctuations due to the crucial role of the Heisenberg interaction $J_t$. Under this condition, the notable \textit{up-down-down (udd) dimerized ground state} can be detected at low enough fields that can be constructed from the eigenvector
\begin{eqnarray}
|{\rm III} \rangle &=& |\!\uparrow \downarrow \downarrow \rangle_{a1,a2,a3} \otimes \frac{1}{\sqrt{2}}
(|\!\uparrow \downarrow \uparrow \rangle - |\!\downarrow \uparrow \uparrow \rangle)_{a4,a5,a6}, \nonumber \\
E_{\rm III} &=& - \frac{3}{4} J_t - \frac{1}{2} J_s - \frac{1}{2} h + \frac{1}{6 \gamma} h.
\label{III}
\end{eqnarray}
The spin alignment inherent to the udd dimerized state unambiguously given by the eigenstate (\ref{III}) is schematically depicted in Fig.~\ref{fig4}c. It can be easily understood from this figure that one out of the three Ising spins belonging to the same spin star (Fig.~\ref{fig2}) is aligned towards the magnetic field, while the other two are oriented in an opposite direction. In addition, there is exactly one field-aligned Heisenberg spin and one singlet dimer per equilateral triangle of the Heisenberg spins (Heisenberg trimer), whereas up-pointing Ising spins are surrounded by singlet dimers and up-pointing Heisenberg spins by down-pointing Ising spins. Altogether, the total magnetization $\frac{1}{2}$ per Heisenberg trimer is just partially compensated by the average magnetization $-\frac{1}{6}$ per Ising spin and hence, the udd dimerized state (\ref{III}) manifests itself through the magnetization plateau at one sixth of the saturation magnetization. If $J_t/J_s \in (\frac{2}{3}, \frac{4}{9} \sqrt{7} - \frac{2}{9})$ [or equivalently $\alpha \in (33.7^{\circ}, 43.6^{\circ})$], the magnetization curves show two intermediate magnetization plateaus at one sixth and one half of the saturation magnetization, which result from two consecutive field-induced transitions between the udd dimerized state, the classical ferrimagnetic state and the saturated paramagnetic state.

Besides, one also encounters an additional magnetization plateau at one third of the saturation magnetization for stronger values of the interaction ratio $J_t/J_s > \frac{4}{9} \sqrt{7} - \frac{2}{9}$, which reflects an appearance of another intriguing \textit{up-up-down (uud) trimerized ground state} stemming from the lowest-energy eigenstate
\begin{eqnarray}
|{\rm IV} \rangle &=& |\!\uparrow \uparrow \downarrow \rangle_{a1,a2,a3} \nonumber \\
&& \otimes \left[\frac{\sin \phi}{\sqrt{2}} (|\!\uparrow \uparrow \downarrow \rangle + |\!\downarrow \uparrow \uparrow \rangle)
- \cos \phi |\!\uparrow \downarrow \uparrow \rangle \right]_{a4,a5,a6}, \nonumber \\
E_{\rm IV} &=& -\frac{1}{2} \sqrt{\left(\frac{J_t}{2} + J_s \right)^2+ 2J_t^2} - \frac{1}{2} h - \frac{1}{6 \gamma} h,
\label{IV}
\end{eqnarray}
where the mixing angle $\phi$ for a quantum superposition of three uud states of the Heisenberg trimers is given by $\phi = \frac{1}{2} \arctan \left( \frac{\sqrt{8} J_t}{J_t + 2J_s} \right)$. A fragment from the spin configuration of the uud trimerized state is schematically illustrated in Fig.~\ref{fig4}d. Apparently, there are two up-pointing and one down-pointing Ising spin per elementary spin star (Fig.~\ref{fig2}) and this non-uniform spin alignment of the Ising spins is also transfered to a quantum superposition of three uud states of the Heisenberg trimers when the highest occurrence probability for a down-pointing spin belongs to the Heisenberg spin attached to the two up-pointing Ising spins. The total magnetization of the uud trimerized state will be determined by the magnetizations $\frac{1}{6}$ per Ising spin and $\frac{1}{2}$ per Heisenberg trimer, which will finally lead to the other magnetization plateau at one third of the saturation magnetization. Bearing this in mind, the magnetization curves exhibit three intermediate magnetization plateaus at one sixth, one third and one half of the saturation magnetization, which are consistent with three consecutive field-induced transitions between the udd dimerized state, the uud trimerized state, the classical ferrimagnetic state and the saturated paramagnetic state whenever $J_t/J_s \in (\frac{4}{9} \sqrt{7} - \frac{2}{9}, \frac{4}{3})$ [$\alpha \in (43.6^{\circ}, 53.1^{\circ})$].

If the interaction ratio exceeds $J_t/J_s > \frac{4}{3}$, there appears another unusual quantum \textit{up-up-up (uuu) dimerized ground state} arising out
from the following eigenstates
\begin{eqnarray}
|{\rm V} \rangle &=& |\!\uparrow \uparrow \uparrow \rangle_{a1,a2,a3} \otimes
\left \{ \begin{array}{l}
\frac{1}{\sqrt{2}} \left(|\!\uparrow \downarrow \uparrow \rangle - |\!\downarrow \uparrow \uparrow \rangle  \right)_{a4,a5,a6}
\\
\frac{1}{\sqrt{2}} \left(|\!\uparrow \uparrow \downarrow \rangle - |\!\uparrow \downarrow \uparrow \rangle  \right)_{a4,a5,a6},
\end{array}\right., \nonumber \\
E_{\rm V} &=& -\frac{3}{4} J_t + \frac{1}{2} J_s - \frac{1}{2} h - \frac{1}{2 \gamma} h.
\label{V}
\end{eqnarray}
It is noteworthy that all the Ising spins are fully aligned in the uuu dimerized state towards the magnetic field, whereas the Heisenberg trimers consist of
one singlet dimer and one field-aligned Heisenberg spin as depicted Fig.~\ref{fig4}e. Hence, it follows that the uuu dimerized state is highly macroscopically
degenerate in contrast to all the previously described ground states, because two linearly independent eigenstates can be constructed for each Heisenberg trimer
from the inner equilateral triangles. Even though the magnetization curves for $J_t/J_s > \frac{4}{3}$ will still contain three intermediate plateaus at one sixth, one third and one half of the saturation magnetization, the last magnetization plateau before the saturation now corresponds to the uuu dimerized state rather than to the classical ferrimagnetic state.

Next, let us compare the magnetization process of the Ising-Heisenberg model (Fig.~\ref{fig3}a) with the relevant numerical data for the analogous but purely quantum Heisenberg model (Fig.~\ref{fig3}b). Except for a few minor differences to be specified in the following, the zero-temperature magnetization curves of the Ising-Heisenberg model and of the quantum Heisenberg model have several important common features. In particular, the plateaus identified at one sixth, one third and one half of the saturation magnetization in the Ising-Heisenberg model can be clearly recognized in the magnetization curve of the Heisenberg model as the largest steps, the smaller ones being very likely finite-size effects for regions in which the magnetization increases smoothly for the infinite system. Of course, it would be necessary to have several sizes at hand and to perform a finite-size analysis to be completely sure about the actual sequence of the plateaus in the Heisenberg model, but the remarkable agreement with the Ising-Heisenberg model makes it likely that all the plateaus of the Ising-Heisenberg case survive. Thus, the only significant difference in the magnetization process of the Ising-Heisenberg and Heisenberg models appears to be the presence of direct magnetization jumps between plateaus in the former classical--quantum Ising-Heisenberg model, while gapless phases in which the magnetization increases smoothly between plateaus are expected to be generically present in the Heisenberg model. For instance, it is quite clear from Fig.~\ref{fig3}b that the magnetization plateau at one half of the saturation magnetization in the parameter region $J_s \gg J_t$ is followed by a smooth continuous increase of the magnetization over a relatively wide field region before reaching the saturation magnetization.

Finally, let us bring some insight into the nature of the gapful phases of the quantum Heisenberg model, which correspond to the most robust magnetization plateaus observable at one sixth, one third and one half of the saturation magnetization. Even though the exact ground states of the Ising-Heisenberg model do not represent the true ground states for the purely quantum Heisenberg model, they might provide at least some useful hint to the true ground states of the quantum Heisenberg model. It should be stressed, however, that the spins from the outer isosceles triangles do not represent an insurmountable barrier for quantum fluctuations
in the Heisenberg model, which may consequently cause a quantum reduction of the magnetization for both kinds of spins from the inner equilateral as well as outer isosceles triangles.

The nature of the ground states is most evident for the two gapped phases which lead to the presence of an intermediate one-half plateau in the magnetization process. The first one-half plateau that emerges in the region with dominant interaction $J_s$ corresponds to a \textit{quantum ferrimagnetic phase}, a quantum analog of the classical ferrimagnetic state shown in Fig.~\ref{fig4}a. Our exact diagonalization data for the nearest-neighbor spin-spin correlation functions of the largest 24-site parallelogram spin cluster indeed provide a strong support of this statement. Considering for instance the particular case with $J_t/J_s = 0$ (i.e. $\alpha = 0^{\circ}$), the correlation functions $\langle \hat{\textbf{S}}_{4i-2} \cdot \hat{\textbf{S}}_{4i-1} \rangle \!=\! 0.23256$ and $\langle \hat{\textbf{S}}_{4i-1} \cdot \hat{\textbf{S}}_{4i} \rangle \!=\! \langle \hat{\textbf{S}}_{4i} \cdot \hat{\textbf{S}}_{4i-2} \rangle \!=\! 0.23385$ imply a strong ferromagnetic correlation between the nearest-neighbor spins from the inner equilateral triangles, see Fig.~\ref{figed}a for the relevant numbering of lattice sites to be congruent modulo 24. By contrast, the correlation functions $\langle \hat{\textbf{S}}_{4i-3} \cdot \hat{\textbf{S}}_{4i} \rangle \!=\! \langle \hat{\textbf{S}}_{4i-3} \cdot \hat{\textbf{S}}_{4i+16} \rangle \!=\! -0.32053$ and $\langle \hat{\textbf{S}}_{4i-3} \cdot \hat{\textbf{S}}_{4i-5} \rangle \!=\! \langle \hat{\textbf{S}}_{4i-3} \cdot \hat{\textbf{S}}_{4i-6} \rangle \!=\! \langle \hat{\textbf{S}}_{4i-3} \cdot \hat{\textbf{S}}_{4i-2} \rangle \!=\! \langle \hat{\textbf{S}}_{4i-3} \cdot \hat{\textbf{S}}_{4i-9} \rangle \!=\! -0.32825$ indicate a relatively strong antiferromagnetic correlation between the nearest-neighbor spins from the inner equilateral and outer isosceles triangles, respectively. On the basis of these results, it could be concluded that the one-half magnetization plateau in the region with dominant interaction $J_s$ indeed corresponds to the quantum ferrimagnetic phase in which both kinds of magnetic moments are subject to a quantum reduction of the magnetization in contrast to the fully saturated magnetic moments of the classical ferrimagnetic state of the Ising-Heisenberg model. Note that the small differences in the relative strength of the nearest-neighbor pair correlation functions can be attributed to the asymmetry of the 24-site parallelogram spin cluster. If the Heisenberg interaction $J_t$ inside the inner equilateral triangles is turned on, the quantum reduction of the magnetization is enhanced due to the spin frustration originating from this interaction term. As a matter of fact, one generally observes a gradual weakening of the correlation functions between the nearest-neighbor spins from the inner equilateral triangles $\langle \hat{\textbf{S}}_{4i-2} \cdot \hat{\textbf{S}}_{4i-1} \rangle \!=\! 0.22449$, $\langle \hat{\textbf{S}}_{4i-1} \cdot \hat{\textbf{S}}_{4i} \rangle \!=\! \langle \hat{\textbf{S}}_{4i} \cdot \hat{\textbf{S}}_{4i-2} \rangle \!=\! 0.22671$, as well as, the correlation functions between the nearest-neighbor spins from the inner equilateral and outer isosceles triangles $\langle \hat{\textbf{S}}_{4i-3} \cdot \hat{\textbf{S}}_{4i} \rangle \!=\! \langle \hat{\textbf{S}}_{4i-3} \cdot \hat{\textbf{S}}_{4i+16} \rangle \!=\! -0.31836$, $\langle \hat{\textbf{S}}_{4i-3} \cdot \hat{\textbf{S}}_{4i-5} \rangle \!=\! \langle \hat{\textbf{S}}_{4i-3} \cdot \hat{\textbf{S}}_{4i-6} \rangle \!=\! -0.32675, \langle \hat{\textbf{S}}_{4i-3} \cdot \hat{\textbf{S}}_{4i-2} \rangle \!=\! \langle \hat{\textbf{S}}_{4i-3} \cdot \hat{\textbf{S}}_{4i-9} \rangle \!=\! -0.32886$  as evidenced by the above results calculated for the particular case with $J_t/J_s = \frac{1}{2}$ (i.e. $\alpha = 26.6^{\circ}$).

Next, let us discuss the nature of the other one-half magnetization plateau, which appears in the region with dominant interaction $J_t$. Let us recall first that the exact ground state of the Ising-Heisenberg model in this parameter range is constituted by the macroscopically degenerate \textit{uuu dimerized state}. The uuu dimerized state includes the fully polarized Ising spins from the outer isosceles triangles, while there is one singlet dimer and one polarized spin per Heisenberg trimer residing on the inner equilateral triangle (see Fig.~\ref{fig4}e). Due to the macroscopic degeneracy of the uuu dimerized state, one should also expect a highly resonating character of the relevant ground-state manifold with the singlet-dimer state being equally distributed over all three bonds of each inner equilateral triangle. Based on these considerations, the correlation function between the nearest-neighbor Heisenberg spins from the inner equilateral triangles should be equal to $\langle \hat{\textbf{S}}_{i} \cdot \hat{\textbf{S}}_{j} \rangle \!=\! -0.25$, whereas the correlation function between the nearest-neighbor Heisenberg and Ising spins from the inner equilateral and outer isosceles triangles should be equal to $\langle \hat{S}_{k}^z \hat{S}_{l}^z \rangle \!=\! \frac{1}{12} \approx 0.08333$. The exact diagonalization data for the nearest-neighbor pair correlation functions bear evidence that the ground state of the pure quantum Heisenberg model is quite reminiscent of the uuu dimerized ground state of the Ising-Heisenberg model. For illustration, let us quote the correlation functions between the nearest-neighbor spins from the inner equilateral triangles of the 24-site parallelogram spin cluster $\langle \hat{\textbf{S}}_{4i-2} \cdot \hat{\textbf{S}}_{4i-1} \rangle \!=\! -0.24500$, $\langle \hat{\textbf{S}}_{4i-1} \cdot \hat{\textbf{S}}_{4i} \rangle \!=\! \langle \hat{\textbf{S}}_{4i} \cdot \hat{\textbf{S}}_{4i-2} \rangle \!=\! -0.24431$ along with the correlation functions between the nearest-neighbor spins from the inner equilateral and outer isosceles triangles $\langle \hat{\textbf{S}}_{4i-3} \cdot \hat{\textbf{S}}_{4i} \rangle \!=\! \langle \hat{\textbf{S}}_{4i-3} \cdot \hat{\textbf{S}}_{4i+16} \rangle \!=\! 0.06370$, $\langle \hat{\textbf{S}}_{4i-3} \cdot \hat{\textbf{S}}_{4i-5} \rangle \!=\! \langle \hat{\textbf{S}}_{4i-3} \cdot \hat{\textbf{S}}_{4i-6} \rangle \!=\! 0.06366, \langle \hat{\textbf{S}}_{4i-3} \cdot \hat{\textbf{S}}_{4i-2} \rangle \!=\! \langle \hat{\textbf{S}}_{4i-3} \cdot \hat{\textbf{S}}_{4i-9} \rangle \!=\! 0.06397$ calculated for the particular case with $J_t/J_s = 4$ (i.e. $\alpha = 76^{\circ}$). It could be easily understood from these results that the true ground state of the Heisenberg model quite closely resembles the uuu dimerized state of the Ising-Heisenberg model, which means that the magnetic moments of the Heisenberg spins from the outer isosceles triangles underlie only a small quantum reduction of the magnetization (they are almost fully saturated) in contrast with the magnetic moments of the Heisenberg spins from the inner equilateral triangles underlying a quantum reduction of the magnetization approximately to one third of the saturation value. As far as the other two gapped phases corresponding to the one-sixth and one-third plateaus are concerned, the relatively strong quantum reduction of the magnetization and the interplay between $J_t \approx J_s$ preclude a simple interpretation of those two gapped phases based on the more subtle udd dimerized and uud trimerized ground states of the Ising-Heisenberg model.

\begin{figure}[t]
\vspace{0.0cm}
\includegraphics[width=10cm]{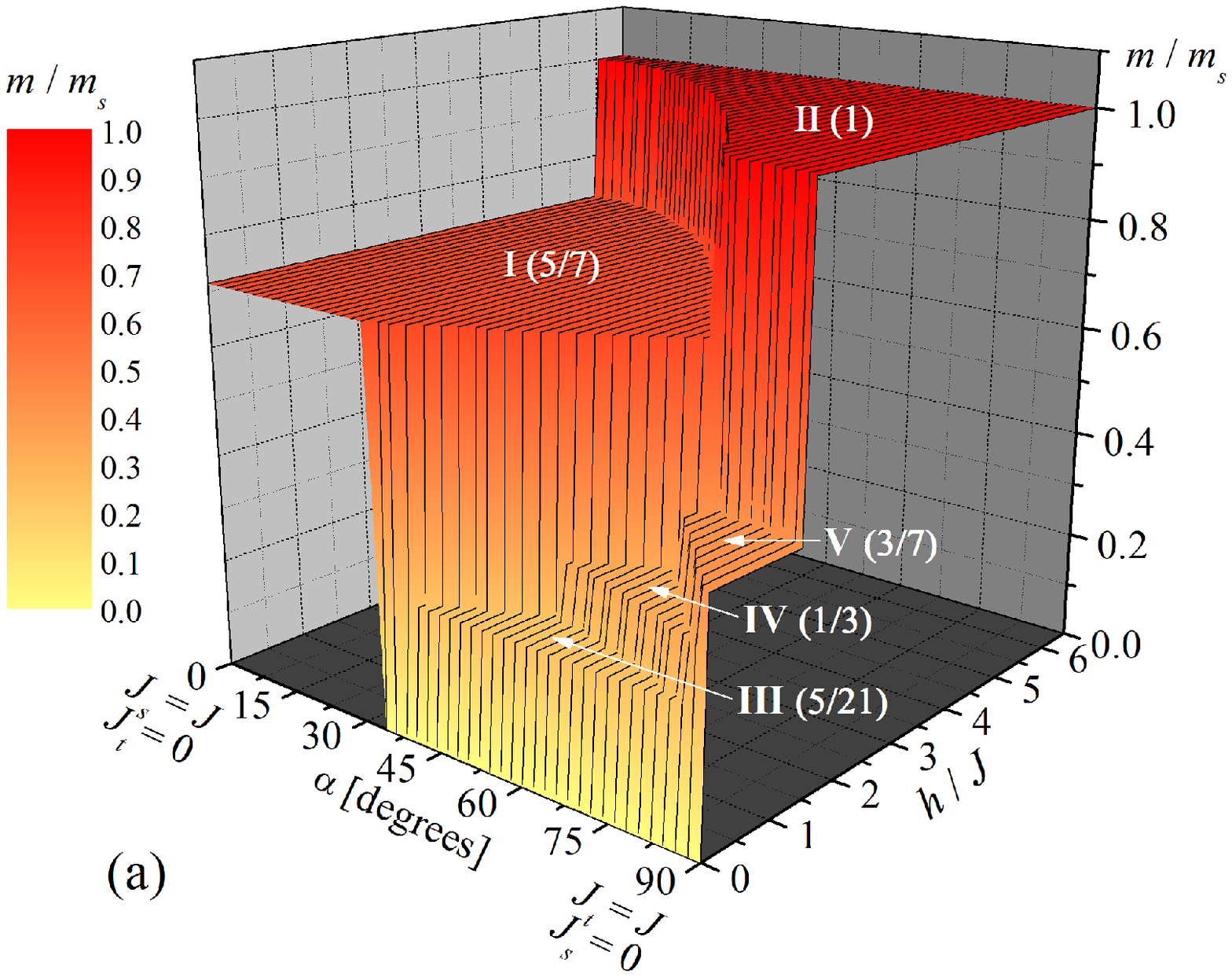}
\includegraphics[width=10cm]{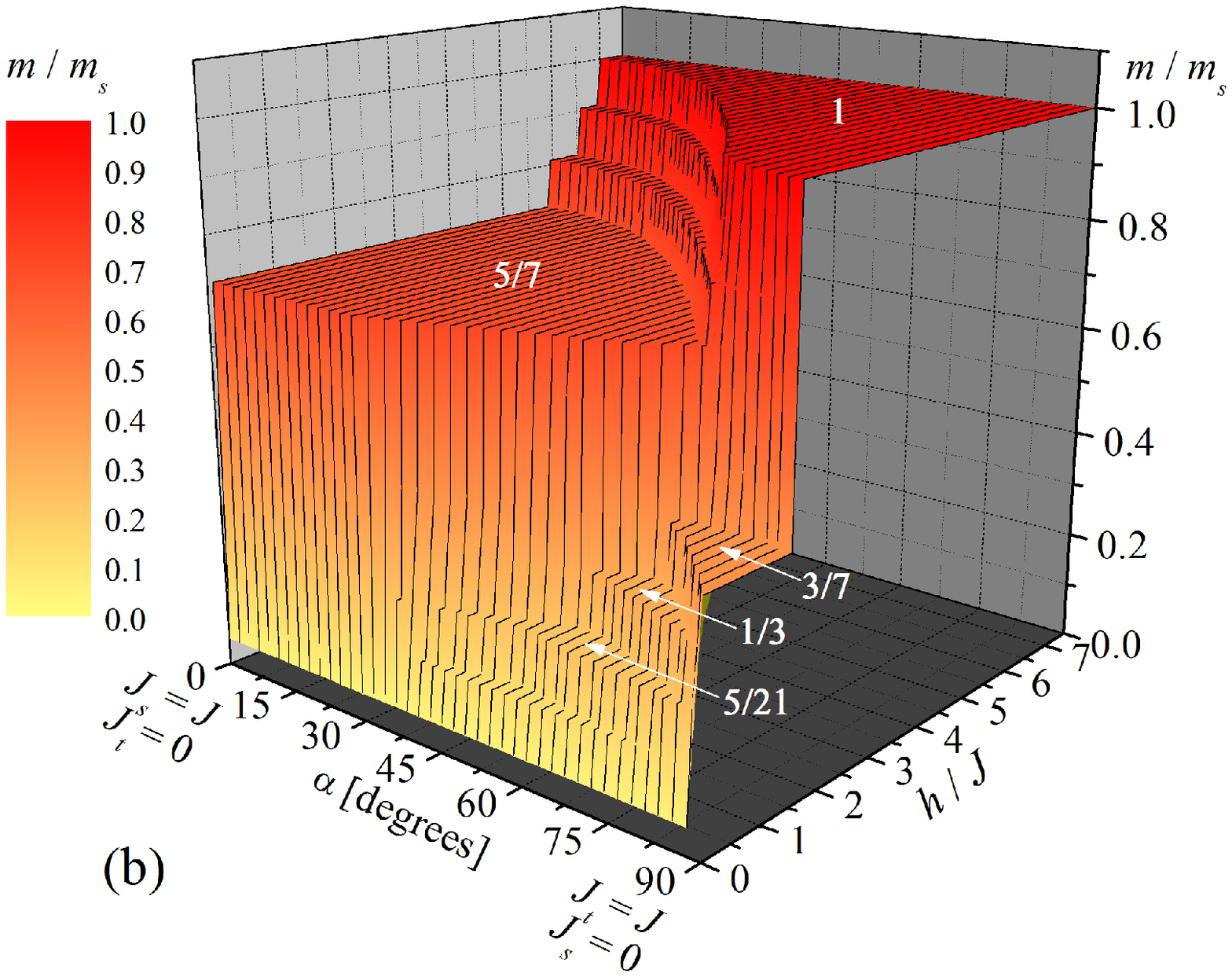}
\vspace{-0.7cm}
\caption{(a) A three-dimensional plot for the zero-temperature magnetization of the spin-$\frac{1}{2}$ Ising-Heisenberg model on the second TIT lattice (Fig.~\ref{fig1}b) as a function of the magnetic field $H/J$ and the parameter $\alpha$; (b) The same three-dimensional plot but for the pure quantum spin-$\frac{1}{2}$ Heisenberg model on the second TIT lattice of 21 sites. Note that since the number of spins is odd, the magnetization starts at 1/2.}
\label{fig5}
\end{figure}

Last, the magnetization process of the spin-$\frac{1}{2}$ Ising-Heisenberg and Heisenberg model on the second TIT lattice (Fig.~\ref{fig1}b) is displayed
in Fig.~\ref{fig5}. It is quite clear that the ground states of the other two Ising-Heisenberg and Heisenberg models are quite analogous with the ones previously described for the first TIT lattice (Fig.~\ref{fig1}a) and hence, there is no need to repeat the comprehensive discussion concerning the nature of possible ground states. It will be sufficient to mention the most significant differences. First, it should be pointed out that the ground states which have exactly the same character will eventually lead to 
the magnetization plateaus at different fractions of the saturation magnetization due to the fact that the number of quantum Heisenberg spins from the inner equilateral triangles is twice as large (the elementary unit cell now consists of six spins from two inner equilateral triangles and one spin from the outer isosceles triangle). Consequently, the spin-$\frac{1}{2}$ Ising-Heisenberg and Heisenberg models on the second TIT lattice exhibit the most robust magnetization plateaus at $\frac{5}{21}$, $\frac{1}{3}$, $\frac{3}{7}$ and $\frac{5}{7}$ of the saturation magnetization, which correspond to the udd dimerized state (\ref{III}), the uud trimerized state (\ref{IV}), the uuu dimerized state (\ref{V}) and the ferrimagnetic state (\ref{I}), respectively. It is worth mentioning that the classical (or quantum) ferrimagnetic state gives rise to a fractional magnetization plateau different from the uuu dimerized state (\ref{V}), which appear at $\frac{5}{7}$ and $\frac{3}{7}$ of the saturation magnetization, respectively. Owing to this fact, the spin-$\frac{1}{2}$ Ising-Heisenberg model on the second TIT lattice shows in a small parameter range $J_t/J_s \in (\frac{19}{12} + \frac{\sqrt{201}}{12}, \frac{10}{3})$  [or $\alpha \in (70.1^{\circ}, 73.3^{\circ})$] a rather spectacular magnetization curve with four different intermediate plateaus, which reflect four consecutive field-induced transitions between the udd dimerized state (III), the uud trimerized state (IV), the uuu dimerized state (V), the classical ferrimagnetic state (I) and the
saturated paramagnetic state (II) in a respective order along ascending field.

\begin{figure}[t]
\vspace{0.0cm}
\includegraphics[width=8.5cm]{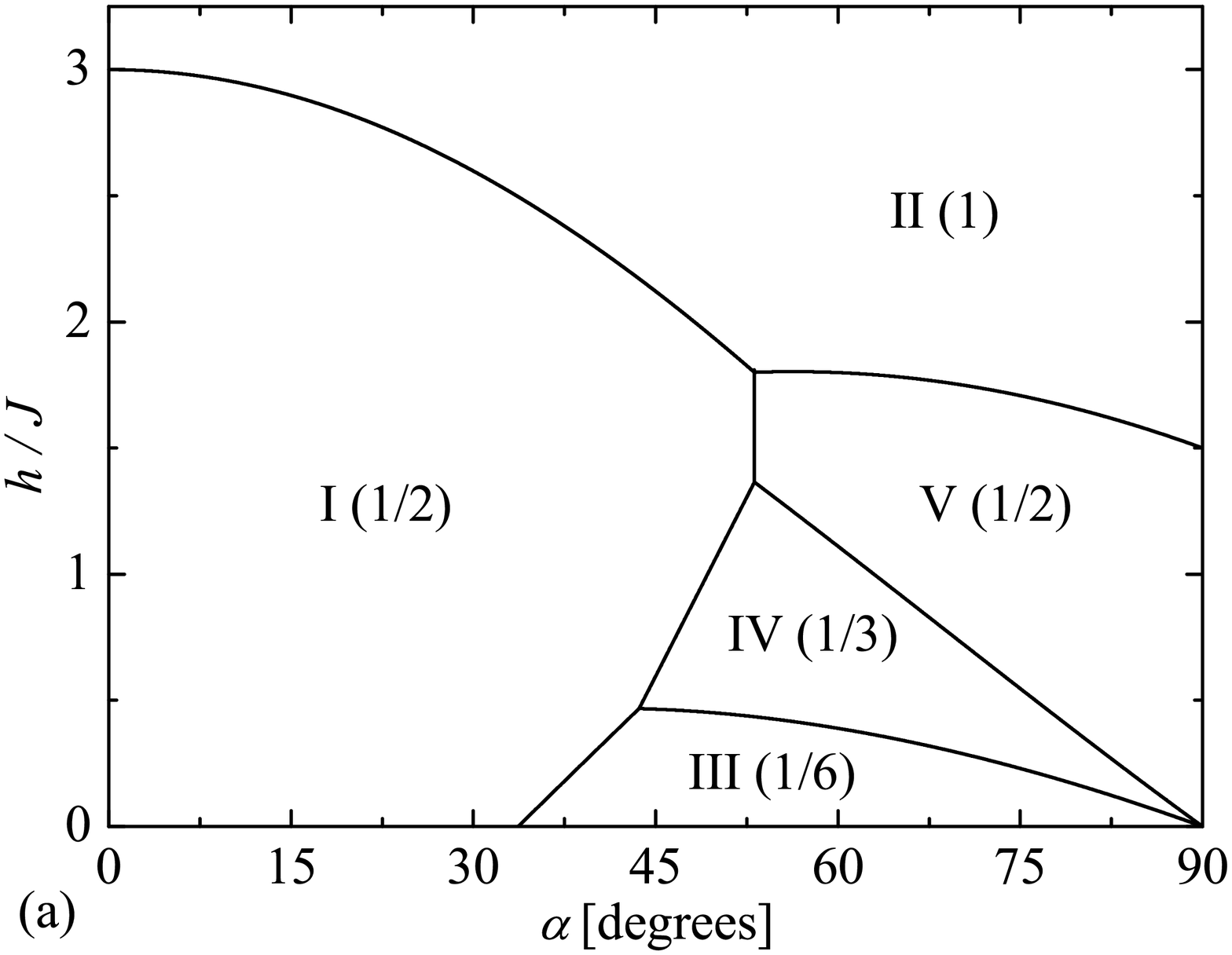}
\includegraphics[width=8.5cm]{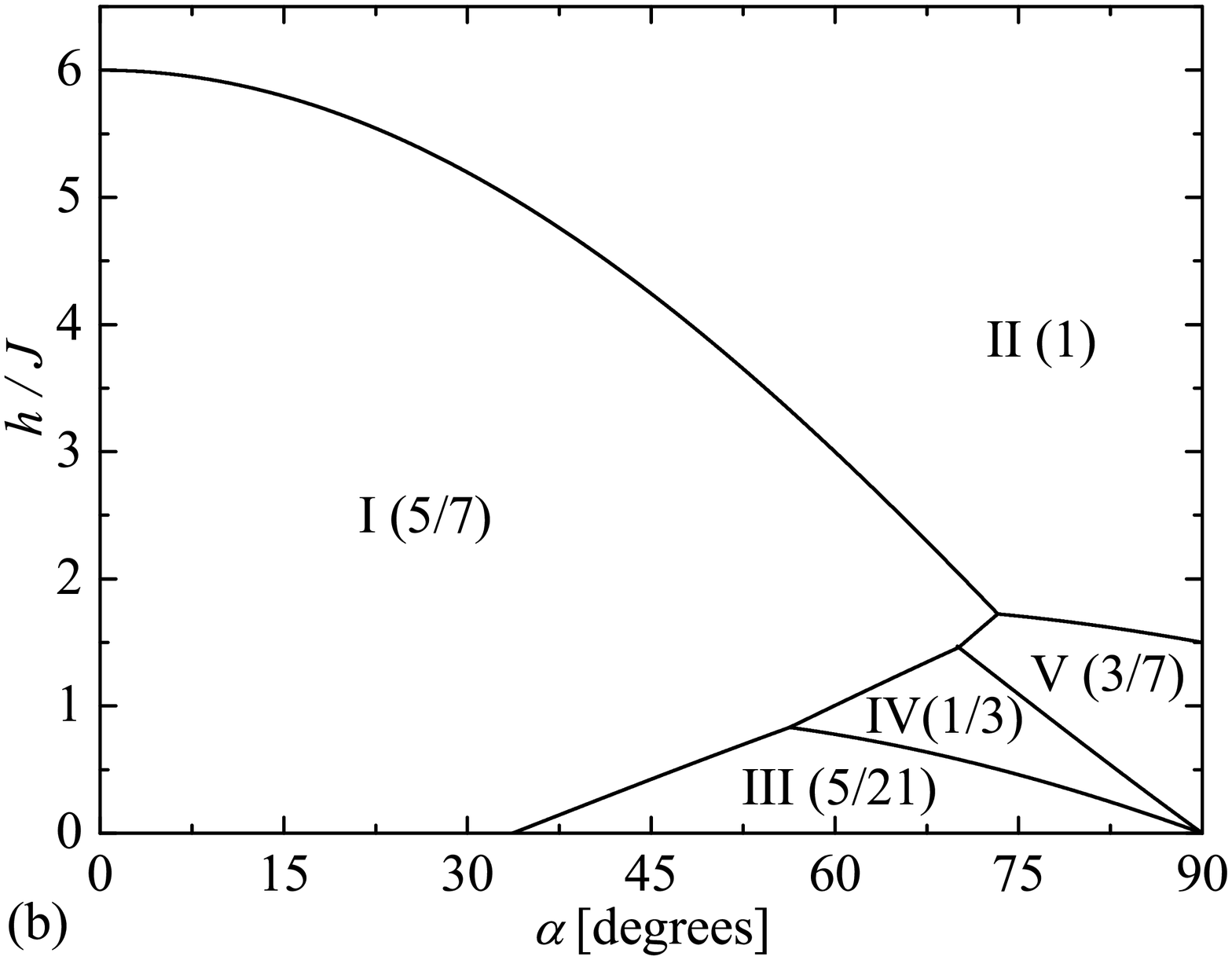}
\vspace{-0.5cm}
\caption{The ground-state phase diagram of the spin-$\frac{1}{2}$ Ising-Heisenberg model on the first (Fig.~\ref{fig6}a) and the second (Fig.~\ref{fig6}b) TIT lattice. The numbers in round brackets refer to the total magnetization normalized with respect to its saturation value.}
\label{fig6}
\end{figure}

To get a global view of the difference between the magnetic behavior of the spin-$\frac{1}{2}$ Ising-Heisenberg model on two considered TIT lattices, the relevant ground-state phase diagrams in the $\alpha - h/J$ plane are displayed in Fig.~\ref{fig6} for comparison. It is quite apparent from this figure that the classical ferrimagnetic state (I) generally extends over a much wider parameter region at the expense of the udd dimerized (III), the uud trimerized (IV), and the uuu dimerized (V) ground states for the Ising-Heisenberg model on the second TIT lattice. In addition, it can be also clearly seen from Fig.~\ref{fig6}b that the magnetization curve of the Ising-Heisenberg model on the second TIT lattice indeed exhibits four consecutive field-induced transitions between the udd dimerized state (\ref{III}), the uud trimerized state (\ref{IV}), the uuu dimerized state (\ref{V}), the classical ferrimagnetic state (\ref{I}) and the saturated paramagnetic state (\ref{II}), which are reflected in the relevant magnetization jumps between four different intermediate plateaus at $\frac{5}{21}$, $\frac{1}{3}$, $\frac{3}{7}$ and $\frac{5}{7}$ of the saturation magnetization if $\alpha \in (70.1^{\circ}, 73.3^{\circ})$. A magnetization process of this type is not present in the Ising-Heisenberg model on the first TIT lattice, which may show at most three different intermediate plateaus before reaching the saturation magnetization (cf. Fig.~\ref{fig6}a with Fig.~\ref{fig6}b).

\section{Concluding remarks}
\label{sec:conclusion}

In the present article, the ground state and zero-temperature magnetization process of the spin-$\frac{1}{2}$ Ising-Heisenberg and Heisenberg model on two geometrically frustrated TIT lattices have been investigated in detail. The magnetization process of the former classical--quantum Ising-Heisenberg model has been rigorously found using the eigenstates of the smallest commuting spin clusters. On the other hand, the magnetization process of the latter purely quantum Heisenberg model has been studied with the help of numerical diagonalizations of finite-size clusters using the Lanczos algorithm. It has been demonstrated that the magnetization curves of the Ising-Heisenberg and Heisenberg models are quite reminiscent even though the exact ground states for the Ising-Heisenberg model do not represent true ground states for the analogous quantum Heisenberg model.

Our exact analytical results for the Ising-Heisenberg model shed light on an extraordinary diversity of the zero-temperature magnetization curves, which may include up to three or four intermediate magnetization plateaus before reaching the saturation magnetization. As a matter of fact, we provide convincing evidence of the existence of three unconventional quantum ground states referred to as the udd dimerized state, the uud trimerized state and the macroscopically degenerate uuu dimerized state in addition to the classical ferrimagnetic and saturated paramagnetic ground states. The most robust magnetization plateaus of the quantum Heisenberg model remarkably coincide with the relevant magnetization plateaus of the Ising-Heisenberg model, which brings some insight into more complex quantum ground states of the Heisenberg model. Until now, the magnetic behaviors of the classical--quantum Ising-Heisenberg model and of the purely quantum Heisenberg model had only been compared for the particular case of the one-dimensional orthogonal-dimer chain.\cite{ohan12} The present study clearly demonstrates that a lot of insight into unusual quantum ground states can be achieved by looking at simpler classical--quantum spin models. 

Even though our analysis was restricted only to the zero-temperature magnetization process under which the most pronounced manifestations of quantum effects should be anticipated, we also expect interesting effects at finite temperatures. Indeed, a discrete lattice symmetry is broken in the ground state of all plateau phases, and we thus expect the system to undergo a phase transition at finite temperature whenever the ground state is in a plateau state, as observed for instance in SrCu$_2$(BO$_3$)$_2$ (see e.g. Ref.~\onlinecite{takigawa_mila}). The determination of the universality class of the transition for a given plateau, which should a priori depend on how the lattice symmetry is broken in this plateau, is a challenging problem that is left for future investigation.

\begin{acknowledgments}
This work was accomplished under the financial support of the Sciex fellowship No.~11.056. J.\v{C}. and J.S. acknowledge the financial support provided by the grant of the Slovak Research and Development Agency under the contract No.~APVV-0132-11, while F.M. and F.M. acknowledge that of the Swiss National Fund and of MaNEP.
\end{acknowledgments}

\end{document}